\documentclass[preprint,12pt]{elsarticle}
\usepackage{amsmath,amssymb,graphicx}
\usepackage{amsfonts,amsthm}
\usepackage{xcolor}
\usepackage{amsmath,amssymb,graphicx,amsthm}
\usepackage{hyperref}

\begin{document}

\begin{frontmatter}

\title{Transition  states   and entangled mass action law}
\author[LeicMath,NNU]{A.N.~Gorban}

\ead{ag153@le.ac.uk}
\ead{https://orcid.org/0000-0001-6224-1430}
\address[LeicMath]{Department of Mathematics, University of Leicester, Leicester, LE1 7RH, UK}
\address[NNU]{Lobachevsky State University Nizhni Novgorod, Russia}

\date{}

%\maketitle

\begin{abstract}
The classical approaches to the derivation of the (generalized) Mass Action Law (MAL) assume that the intermediate transition state  (i) has short life time and (ii) is in   partial equilibrium with the initial reagents of the elementary reaction. The partial equilibrium assumption (ii) means that the reverse decomposition of the intermediates is much faster than its transition through other channels to the products. In this work we demonstrate how avoiding  this partial equilibrium assumption modifies the reaction rates. This kinetic revision of transition state theory results in an effective `entanglement' of reaction rates, which become linear combinations of different MAL expressions.
\end{abstract}

\end{frontmatter}

%\section{Introduction}

\section{Introduction: Backgrounds of generalized mass action law}

\subsection{ The basic assumptions}

The Mass Action Law (MAL) for chemical kinetics was postulated by Guldberg and Waage first for equilibrium (in 1864), and then for dynamics  (in 1879). Boltzmann used the analogue of the dynamical MAL ({\em Stosszahlansatz}) for collision in gases in  1872 and obtained the gas kinetics equation (Boltzmann's equation).
The physical background of MAL for complex chemical reactions was clarified later, in particular in transition state or activated complex theories  published in 1935 (see review in \cite{Laidler1983}). From the kinetic point of view, the assumptions that lead to MAL can be summarized as follows:
\begin{enumerate}
\item A complex reaction is a combination of elementary reactions;
\item Elementary reactions pass through  intermediate states;
\item Intermediate states have a short lifetime and are present in very small concentrations compared to the main reagents;
\item There is a `fast equilibrium' between the input reagents and intermediate states for each elementary reaction and these fast partial equilibria can be described thermodynamically, by the conditional minimization of the free energy.
\end{enumerate}

\begin{figure*}
\centering{
\includegraphics[height=0.26\textwidth]{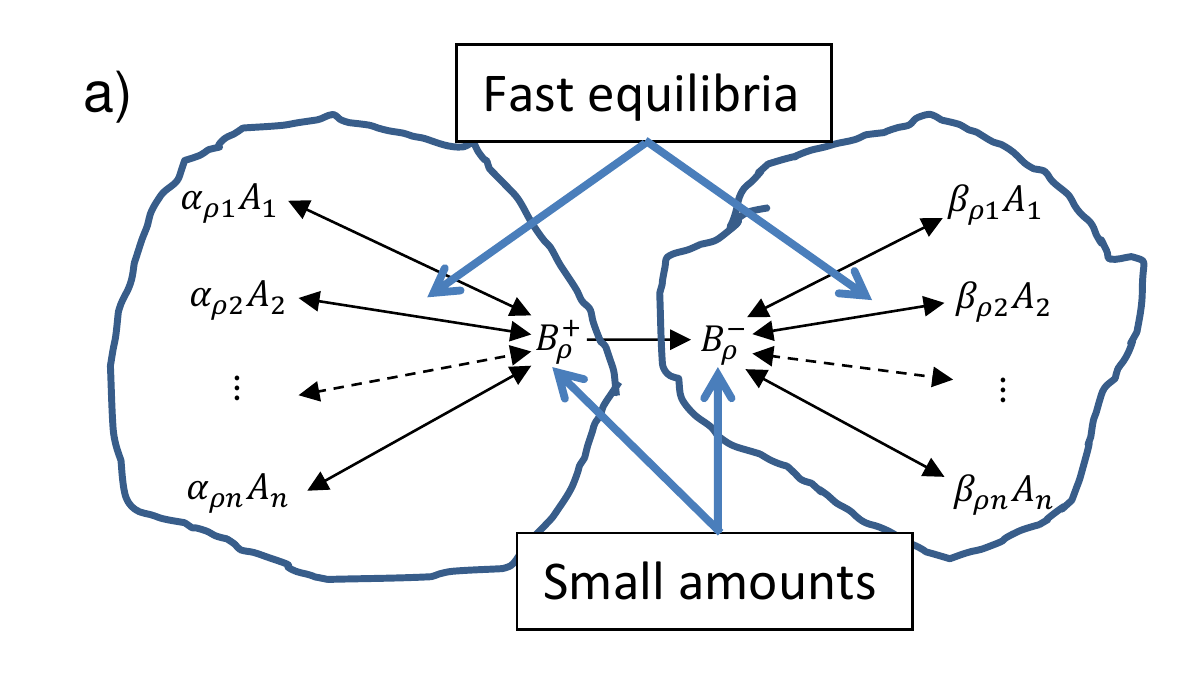}
\includegraphics[height=0.26\textwidth]{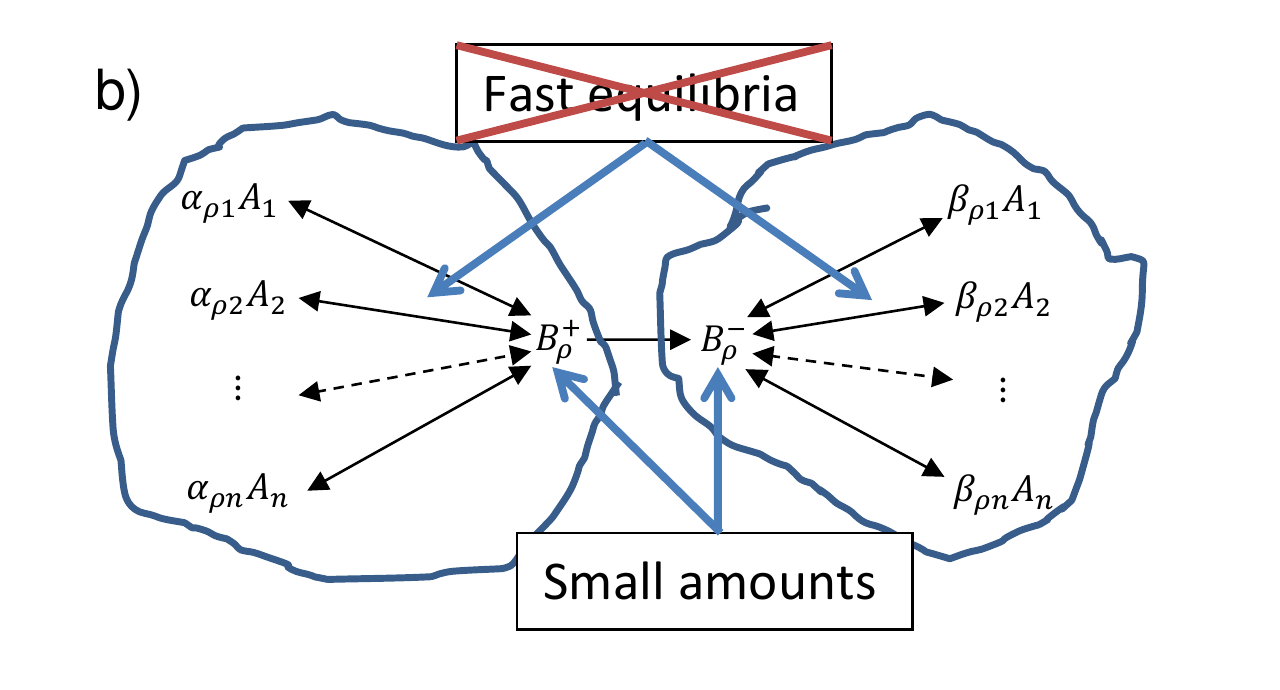}}
\caption{\label{StueckAsympt} Schematic representation of the Michaelis--Menten--Stueckelberg  (a), and Briggs--Haldane  (b) asymptotic assumptions: an elementary process $\sum \alpha_{\rho i}A_i \to \sum \beta_{\rho i}A_i$ goes through intermediate compounds $B_{\rho}^{\pm}$. The fast equilibria in the Michaelis--Menten case (a),
$\sum \alpha_{\rho i}A_i\rightleftharpoons B_{\rho}^+$ and $\sum \beta_{\rho i}A_i\rightleftharpoons B_{\rho}^-$,
can be described by conditional minima of free energy. The concentrations of $B_{\rho}^{\pm}$ are small and the transitions
between them obey   linear kinetic equation (Markov chains).}
\end{figure*}

The kinetic assumptions used in transition state theory were introduced much earlier. In 1913, Michaelis and Menten demonstrated  for an enzyme reaction that under these assumptions the overall reaction follows  MAL \cite{MichaelisMenten1913}. They used the term `compounds' for the active intermediates.  In 1952, Stueckelberg \cite{Stueckelberg1952} used the same assumptions for the derivation of the Boltzmann equation from the Markovian microkinetics and demonstrated how the semidetailed balance conditions (called also the cyclic balance conditions  \cite{Watanabe1955} or the complex balance conditions \cite{HornJackson1972,Feinberg1972}) arose for the cases when  the classical microreversibility conditions failed and the principle of detailed balance did not work. The semidetailed balance was invented by Boltzmann \cite{Boltzmann1887} as an answer to Lorentz critique of microreversibility. Boltzmann's proof of sufficiency of semidetailed balance for $H$-theorem was straightforward. The  assumptions 1--4 are illustrated in  Fig.~\ref{StueckAsympt}a according to Michaelis--Menten and Stueckelberg works.

The proof that the generalized MAL and semidetailed balance should hold for macroscopic kinetics is not so obvious.  Watanabe \cite{Watanabe1955} found that the Stueckelbeck work has ``lack of proof'' and declared direct connection between this condition and ergodicity of microscopic models. Detailed analysis of this asymptotic (Fig.~\ref{StueckAsympt}a)  was performed in 2011,  the general Michaelis--Menten--Stueckelberg (MMS) theorem was formulated and proved for finite-dimensional systems \cite{GorbanShahzad2011}, and extended further for general  nonlinear Markov processes \cite{GorbanKol2015}. The physical assumptions that lead to generalized MAL with detailed or semidetailed (cyclic) balance were analyzed in \cite{Gorban2014}.

Assumption 3 about the short lifetime of transition states underlies the approach to the microscopic theory of reaction rates. This constitutes the difference between the reagents and intermediates. Assumption 4 adds an additional time separation: equilibration between the  input reagents and the intermediate states for every elementary reaction is much faster than other transitions of the intermediates. This traditional hypothesis simplifies the use of microscopic quantum or semiclassical calculations of transition state dynamics in kinetics, but also attracts criticism. Recently, the ideas of a quasi-stationary state and partial equilibrium in the theory of transition states have been analyzed and compared `from scratch' on a simple  example \cite{Perez2017}. This analysis led to a quasi steady state expression for the rate constant of an elementary reaction that differs only by a simple factor from the value given by classical theory. The result was compatible with chemical thermodynamics.

\subsection{ A popular `toy example' of the single transition state  paradox}

It was clearly shown that the popular `demo version' of combining the results of the transition state theory with kinetics (one intermediate compound in partial equilibrium with the reactants, and forward and reverse reactions go through the same intermediate) contradicts chemical thermodynamics. Indeed, consider an elementary reaction
\begin{equation}\label{simplestMM}
A+B \rightleftharpoons [A-B] \rightleftharpoons  P+Q,
\end{equation}
where  $[A-B]$ is the transition state (or better to say `activated complex' because the term `transition state' is often used for the dividing surface that separates the input reagent side from the product side in the state space). The partial equilibrium assumption requires equilibration of both transitions, $A+B \rightleftharpoons [A-B]$ and $[A-B] \rightleftharpoons  P+Q $. There is no essential difference between them, since the reverse reaction proceeds through the same transition state. Straightforward elementary algebra demonstrates that this equilibration implies the equilibrium of the brutto-reaction $A+B \rightleftharpoons  P+Q$, which should be a thermodynamic equilibrium between $A$, $B$ and $P$, $Q$ \cite{Perez2017}: Due to equilibration $A+B \rightleftharpoons [A-B]$,  the concentration $[A-B]$ is a function of the  concentrations of $A$ and $B$, and by equilibration $[A-B] \rightleftharpoons  P+Q $, it is  a function of the concentrations of $P$ and $Q$. Values of these two functions coincide. That gives us the thermodynamic equilibrium condition.  Therefore, the partial equilibrium assumption for  transitions (\ref{simplestMM}) cannot be used for kinetics if we respect thermodynamics. Let us call this contradiction the {\em single transition state  paradox}.

{ 
The general version of this paradox can be represented as follows.
 According to  the standard   basic assumption 4, the transition state is in quasiequailibrium with the reagents, and we can apply the transition state theory and find the reaction rate.  The separation of components into `reagents' and `products' for a reversible reaction is a matter of convention, and the situation is symmetrical with respect to renaming of `reagents' into `products' and vice versa. Thus, we can assume quasiequilibrium between the reagents and the products and find the reverse reaction rate. But if we assume these two quasiequlibria together, then a simple calculation shows that this is an equilibrium of the brutto-reaction and, in particular, the rates of the direct and reverse reactions coincide.  }

This paradox can be resolved by the classical idea: the activated complex should be represented by several macroscopic compounds. This approach is even older than the transition state theory. Already in 1913, Michaelis and Menten \cite{MichaelisMenten1913} considered two intermediates:
 \begin{equation}\label{naturalMM}
A+B \rightleftharpoons [A-B]  \rightleftharpoons [P-Q] \rightleftharpoons  P+Q.
\end{equation}
Here we employ notations from \cite{Perez2017} and use $ [A-B] $ and $[P-Q] $ instead of $B^{\pm}$ (see Fig.~\ref{StueckAsympt}).
According to the Michaelis and Menten assumption, $[A-B]$ is in  partial equilibrium with $ A$ and $B$, whereas $[P-Q]$ is in   partial equilibrium with $P$ and $Q$, and the paradox of single intermediate disappears. We can consider the pair $ [A-B]  \rightleftharpoons [P-Q]$ as a model of the microscopic dynamics of an activated complex. It is quite natural that a microscopic  activated complex with a multidimensional space of internal coordinates should be represented by several macroscopic states. Stueckelberg \cite{Stueckelberg1952} also used the two-intermediate representations of collisions for the Boltzmann equation  and assumed Markovian transitions between these intermediates.

\begin{figure*}
\centering{
\includegraphics[height=0.4\textwidth]{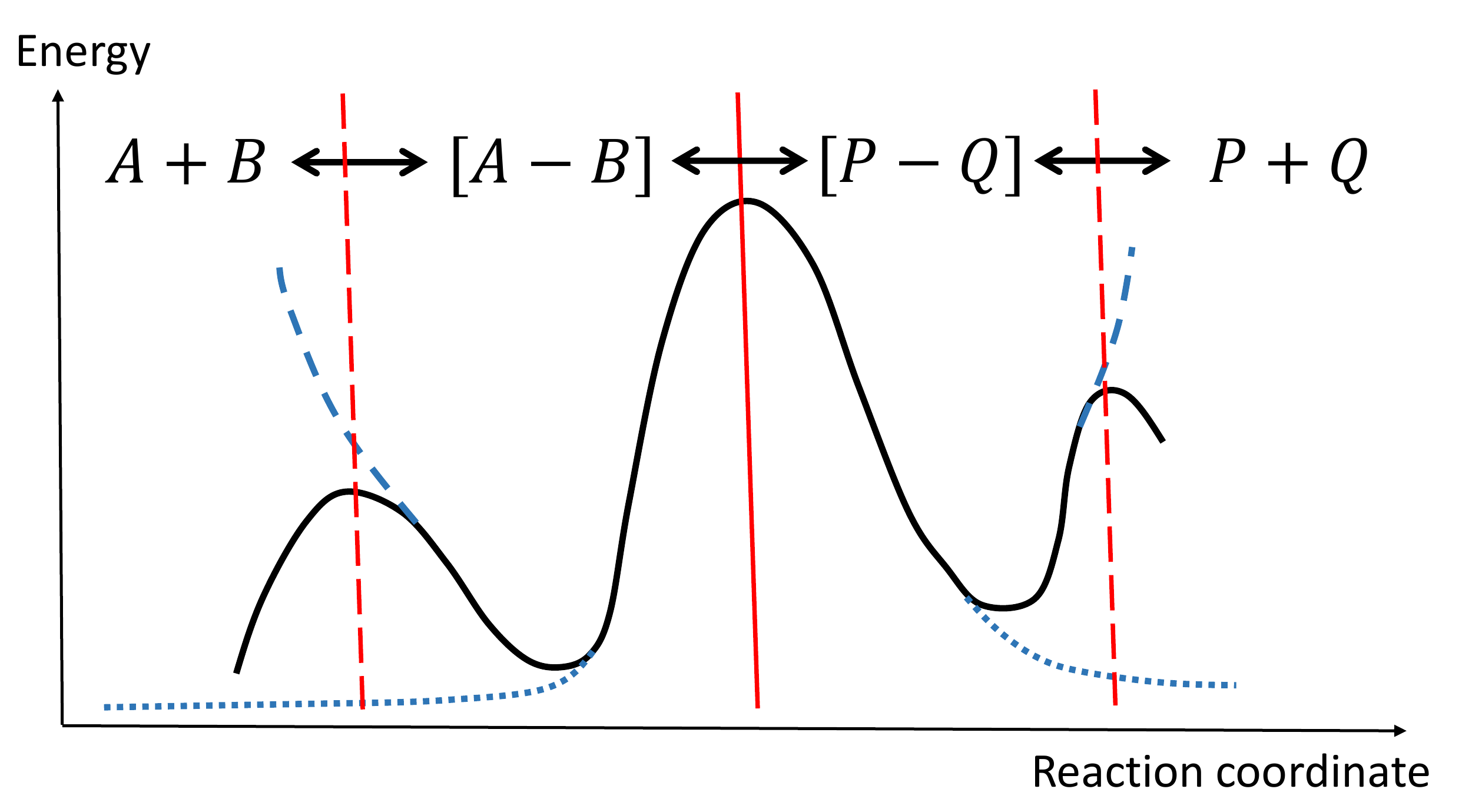}}
\caption{\label{PotentialWellCompounds}Schematic illustration of a two-intermediate representation of the activated complex. The intermediates $[A-B]$ and $[P-Q]$ are metastable states
in the energetic landscape (black solid line). The vertical solid line (red) corresponds to the classical transition state and separates the input equilibrated system $A+B \rightleftharpoons [A-B]$ from the output systems $[P-Q] \rightleftharpoons  P+Q$. The vertical long dashed lines (red) separate   $A+B$ from $[A-B]$ (left) and $[PQ]$ from $ P+Q$ (right). The auxiliary potential wells  (dashed lines, blue) can be used for artificial stabilization of the compounds for computations. There is also the possibility that the energy landscape is strictly unimodal (dashed lines, blue). (Colors online.)}
\end{figure*}

A two-intermediate model of the activated complex is illustrated in Fig.~\ref{PotentialWellCompounds}. There are four ranges of reaction coordinate values. They correspond to $A+B$, $[A-B]$, $[P-Q]$, and $P+Q$. In the MMS assumptions, transitions between $A+B$ and $[A-B]$ are regulated by the partial equilibrium approximation: The concentration of $[A-B]$ follows the concentrations of $A$ and $B$, and the reaction $A+B \rightleftharpoons [A-B]$ is always in equilibrium. The same is true for the transitions $[P-Q] \rightleftharpoons  P+Q$. The transition rate $[A-B]  \rightleftharpoons [P-Q]$ is to be estimated by the transition state theory. It defines the flux between the total masses $A$, $B$, $[A-B]$ and $P$, $Q$, $[P-Q]$. If we assume, in addition, that the intermediates are present in small concentrations (that is, their life time is small), then  the transition rate $[A-B]  \rightleftharpoons [P-Q]$ gives the brutto-reaction rate $A+B  \rightleftharpoons  P+Q$.

\subsection{  Several intermediates and Jump Markov model}

In more general situations, the key steps remain the same: classically define the surface dividing incoming reagents and outgoing products, and calculate the velocity constant as a  phase space  flow through this surface. The classical transition state theory  takes into account the dynamics only near the transition state, which, according to the standard definition, is the dividing  surface separating reactants and products. This assumption persists in many generalized versions of the transitions state theory, in particular,  in  variational transition state theory \cite{Bao2017} and in stochastic transition state theory \cite{Pollak2018}.  Many amendments were made to the classical theory. Consideration of quantum tunneling effects led to a deformed transition state theory and modification of the Arrhenius reaction rate form \cite{Carvalho2017}. It was found that the definition of general dividing surfaces in the quantum theory of the  transition state is fundamentally non-unique \cite{JangVoth2017}.

One of the basic assumptions of  transient state theory is that when the trajectory crosses the dividing surface separating the input reactants from the product, no further recrossing occur. But for the classical definition of a dividing surface, this is not the case: the trajectories can enter the product side of the space, and then return back to the  side of the input reagents. This recrossing  leads to an effective `swelling' of the dividing surface. There are many attempts to improve computational approaches and correct theoretical expressions for reaction rates taking into account the possibility of recrossing \cite{Sharia2016}.

A multidimensional activated complex requires a more detailed partitioning of the state space to create an adequate kinetic Markov model (Master equation) of the reaction. This partitioning may be  based on local energy minima and corresponding metastable states \cite{Gesu2017}. Another idea is to evaluate the residence time in subdomains and define the partitioning based on this time. That is, in addition to metastable states, we can consider long-term transient regimes \cite{GorbanSlow,Kurth2017,GorbanSing2020} as elements of the state space of Markov kinetics. This sort of metastability may be related not to the energetic barriers but to the entropic barriers. According to \cite{Gesu2017}, the entropic barriers  appear when the process takes a long time to
leave the subdomain $S$ because `the exit doors from $S$ are very narrow'.  In chemistry, entropic barriers may be related to steric constraints. In general, metastability of a state is is defined by the ratio of the expected exit time to the expected local equilibration time. If this ratio is much greater than one, then the subdomain can be considered as a metastable state.

In this `Jump Markov' model approach,  instead of the MMS two-intermediate representation shown in Fig.~\ref{StueckAsympt}, a more general network of intermediates and first order transitions  between them appears. However, the applications of this approach to the  derivation of macroscopic kinetic equations is not yet freed from the assumption of partial equilibrium between reactants and intermediates. The quasi steady state approach sketched in  \cite{Perez2017} followed the Briggs-Haldane approximation \cite{BriggsHaldane1925} based on the assumption of small amounts of intermediate (or, specifically,  small enzyme concentrations). This is fine but the Briggs-Haldane approximation uses the dynamic form of MAL for $A+B \rightleftharpoons [A-B]$,  for both direct and reverse  elementary reactions. The linear kinetics of the reverse reaction  $A+B \leftarrow [A-B]$ follows from the smallness of the concentration $[A-B]$, while the bilinear reaction rate of the direct reaction $A+B \to [A-B]$ is an additional assumption. Starting with  \cite{MichaelisMenten1913},  the assumption of a fast equilibrium (between input reagents and intermediate products) is used to obtain the dynamic MAL or its generalizations. If we weaken this assumption, we will need a universal approximation close to partial equilibrium. Linearization near partial equilibrium seems to be a suitable and fairly universal choice. Utilization of the Briggs-Haldane  quasi steady state approximation produces the transmission coefficient that modifies the transition theory reaction rate (see \cite{Perez2017}).
Our results give the generalized matrix  transmission coefficient that we call the entanglement matrix (see Eqs.~(\ref{result1}), (\ref{result2}), and (\ref{qssreactionrate} below).

In this paper we do not try to modify the transition state theory. In the following, the microscopic theory is just a `black box'  that provides us with the Markov kinetics of   intermediates. For us, the main question is: how to  correctly incorporate this black box  in  the basic schemes of thermodynamics and kinetics. We demonstrate the possible development of the reaction rate theory without assumption 4. No super-fast equilibration is assumed and all  transitions of intermediates can have a comparable characteristic time.  Already in the first approximation, the reaction rate theory based on assumptions 1-3 produces the `entangled' MAL: the reaction rates for elementary reactions become combinations of the MAL rates of various elementary reactions.

\section{The basic formalism}

We use the following widely used notations borrowed from chemical kinetics \cite{GorbanShahzad2011,MarinYab2019}. The list of
the components (reagents) $A_i$ is given. The mechanism of reaction is the list
of the elementary reactions. Each reaction is represented by the stoichiometric
equation:
\begin{equation}\label{elementary reaction}
\sum_i\alpha_{\rho i}A_i \to \sum_i \beta_{\rho i} A_i \, ,
\end{equation}
where $\rho$ is the number of the elementary reaction qand the stiochiomentric coefficients $\alpha_{\rho i}$, $\beta_{\rho i}$ are non-negative numbers; $\alpha_{\rho }$, $\beta_{\rho }$ are vectors wit components $\alpha_{\rho i}$, $\beta_{\rho i}$, correspondingly. The concentration of the component $A_i$ is $c_i$, the vector of concentrations is $c$. The kinetic equations for (\ref{elementary reaction}) under isochoric isothermal conditions are (with obvious modifications for other conditions \cite{GorbanShahzad2011}):
\begin{equation}\label{kinur}
\frac{dc}{dt}=\sum_{\rho}(\beta_{\rho }-\alpha_{\rho })r_{\rho},
\end{equation}
where $r_{\rho}\geq 0$ is the reaction rate.
The linear combinations $\sum_i\alpha_{\rho i}A_i$ and $\sum_i \beta_{\rho i} A_i$ are the {\em complexes}. Each complex can participate in various reactions. The list of coefficient vectors for complexes is $\{\nu_j\}$. For each complex $\sum_i \nu_{ji} A_i$ from the reaction mechanism we introduce an intermediate state, a {\em compound} (in Michaelis--Menten terminology) $B_j$. Each
elementary reaction is represented in the form of the ``$2n$-tail
scheme''   with two intermediate compounds:
\begin{equation}\label{stoichiometricequationcompaundI}
\sum_i\alpha_{\rho i}A_i \rightleftharpoons B_{\rho}^+ \to
B_{\rho}^- \rightleftharpoons \sum_i \beta_{\rho i} A_i
\, ,
\end{equation}
where the compound $B_{\rho}^+$ corresponds to the input complex $\sum_i\alpha_{\rho i}A_i$ of  (\ref{elementary reaction}), and $B_{\rho}^-$ corresponds to the output complex  $\sum_i \beta_{\rho i} A_i$. The concentration of the compound $B_j$ is $\varsigma_j$, the vector of these concentrations is $\varsigma$. Thermodynamical properties of the mixture is described by the free energy density (under the condition of small concentrations of compounds, $\sum_i c_i\gg \sum_j \varsigma_j$, the  free energy of the small admixture has the perfect form):
\begin{equation}\label{FreeEn2}
f(c,\varsigma,T)=f_A(c,T)+RT \sum_{j=1}^q \varsigma_j \left(\ln
\left(\frac{\varsigma_j}{\varsigma_j^*(c,T)}\right)-1\right) \
\end{equation}
We assume that the standard equilibrium concentrations
$\varsigma_j^*$ are much smaller than the concentrations of
$A_i$, $\sum_i c_i\gg \sum_j \varsigma_j^*$.
The ``fast equilibrium'' reagents-compounds for a complex $\sum_i \nu_{ji} A_i$  is described by the conditional minimisation of $f(c,\varsigma,T)$ along the straight line
$c=c^{\rm in}-\xi \nu_{j}$, $\varsigma_j=\varsigma_j^{\rm in}+\xi \nu_{j}$, where superscript `in' is used for the initial point for minimisation, and $\xi$ is the coordinate along the line.

For the kinetics of compounds transformations, the  smallness of concentrations of compounds leads to the   linear (Markov) kinetics, where the rate constant $\kappa_{lj}$ of transitions $B_j \to B_l$ (or $\kappa_{l\leftarrow j}$ if we need to stress the direction of transition) can depend on concentrations $c$ and temperature $T$.

According to the basic MMS asymptotics, there are two small parameters, $\delta$  that evaluates the relative deviation of $\varsigma$ from their conditional equilibrium approximation, and $\varepsilon$ that evaluates the relative smallness of  $\varsigma$ comparing to $c$ \cite{GorbanShahzad2011}. The compound rate constants should be rescaled when  $\varepsilon \to 0$ as $\frac{1}{\varepsilon}\kappa_{lj}$.
In the asymptotic $\delta, \varepsilon \to
0$, $\delta, \varepsilon > 0$ kinetics of $A_i$ may be
described by the reaction mechanism (\ref{elementary reaction})
with the reaction rates
\begin{equation}\label{GMAL}
r_{\rho}=\varphi_{\rho}\exp\left( \frac{\sum_i \alpha_{\rho i} {\mu}_i}{RT}\right)
\end{equation}
where where $\mu_i=\frac{\partial f(c,T)}{\partial c_i}$ is the chemical
potential of $A_i$ and the kinetic factors $\varphi_{\rho}$ satisfy the  condition:
\begin{equation}\label{complexbalance}
\sum_{\rho, \,\alpha_{\rho}=y} \varphi_{\rho}\equiv
\sum_{\rho, \,\beta_{\rho}=y} \varphi_{\rho}
\end{equation}
for any vector $y$ from the set of all vectors
$\{\alpha_{\rho}, \beta_{\rho}\}$. This statement includes the
generalized mass action law for $r_{\rho}$ (\ref{GMAL}) and the semidetailed  balance identity (\ref{complexbalance})
for kinetic factors that is sufficient for the entropy growth. If the Markov chain of compound kinetics is reversible (i.e. satisfies the principle of detailed balance), then for the generalized MAL the priciple of detailed balance holds in the form: 
\begin{equation}\label{DetBal}
\varphi_{\rho}^+\equiv \varphi_{\rho}^-.
\end{equation}

{ 
In terminology of  nonequilibrium thermodynamics (see, for example, chapter 4 in  \cite{Grmela2018}) we can refer to kinetic equations (\ref{kinur}) as a local conservation law; the stoichiometric matrix plays the role of the gradient in the classical local conservation laws and (\ref{GMAL}) is a (generalized) MAL constitutive relation. Nevertheless, despite possible generalizations, we do not use such a general terminology and corresponding notations and remain at the level of chemical kinetics \cite{MarinYab2019}.

The form of generalized MAL may need additional comments. The {\em Bolzmann factor} (the exponential of linear combinations of chemical potentials) is the product of so-called activities:
$$\exp\left( \frac{\sum_i \alpha_{\rho i} {\mu}_i}{RT}\right)=\prod a_i^{\alpha_{\rho i} },$$
where $a_i$ are activities: $$ a_i=\exp\left(\frac{{\mu}_i}{RT}\right).$$

For perfect systems, for example, $\mu_i=\ln(c_i/c_i^*)$, where $c_i^*$ are concentrations in standard equilibrium, and $a_i=c_i/c_i^*$. in this (most common) case, the Boltzmann factor in generalized MAL formula (\ref{GMAL}) turns into classical MAL form, the product of concentrations.

Nevertheless, the kinetic factor $\varphi_{\rho}\geq 0$ can violate the widely spread intuition that the MAL reaction rate  is the power law with a multiplier. power law is a function of the concentrations or the activities of the reactants, and the multiplier is a constant independent on the concentrations or the activities of any reactant. In  (\ref{GMAL}),   the kinetic factors  $\varphi_{\rho}$ may depend on concentrations. The conditions on these factors are either detailed balance  (\ref{DetBal}) for systems with microrevesibility, or more general complex balance (\ref{complexbalance}) for systems with possible violations of microreversibility. Appearance of such general factors is a well-known in kinetics of generalized MAL (see, for example, Marselin--de Donder kinetics and is generalizations  \cite{Feinberg1972MDD, BykGorYab1982MDD}).

A simple physical example can add clarity. In gas reactions, some collisions need a `third particle' (or `third body') that does not change in reaction but is needed for energy or momentum balance. Usual notation for this third particle is M. For example, the following reactions exist in the standard mechanism of hydrogen burning \cite{Conaireatal2004}:
${\rm H}_2+ {\rm M} \rightleftharpoons 2 {\rm H} +{\rm M}$, ${\rm H} + {\rm OH}+ {\rm M}\rightleftharpoons{\rm H}_2{\rm O}+{\rm M}$, or ${\rm H} +{\rm O}_2+{\rm M}\rightleftharpoons {\rm HO}_2+ {\rm M}$.

For different third particles the reaction rates may differ, therefore each such stoichiometric equation denotes a set of reactions with different M. Nevertheless, for the reaction rate of such a multiplied reaction the generalized MAL holds. Formally, it relates the generalized reaction without M:  the Boltzmann factor corresponds to the  MAL calculated  for components without M, and the kinetic factor depends on concentrations of all M's. In the simplest case, for perfect gases $\varphi_{\rho}$  are linear functions of concentrations of third particles with positive coefficients. Thus, we can say that dependence of kinetic factors on the concentrations in this example describes the impact of the environment on the reaction rate.

 For reactions in condensed matter or on surfaces, the impact of the environment on the reaction rate may be more sophisticated. It does not effect the thermodynamic equilibrium governed by the  thermodynamic potentials, but modifies the kinetic factors.

Dependence of  $\varphi_{\rho}$ on concentrations  in kinetic equations can have even  simpler reasons. For example, if we consider a reaction in an isolated system, then the temperature can be expressed as a function of concentrations and (constant) internal energy. In such a situation, $\varphi_{\rho}$ in kinetic equations are functions of concentrations even if initially they depended on temperature only \cite{BykGorYab1982MDD}.}

\section{The problem of fast non-equilibrium intermediates}

\begin{figure}
\centering{
\includegraphics[width=0.7\textwidth]{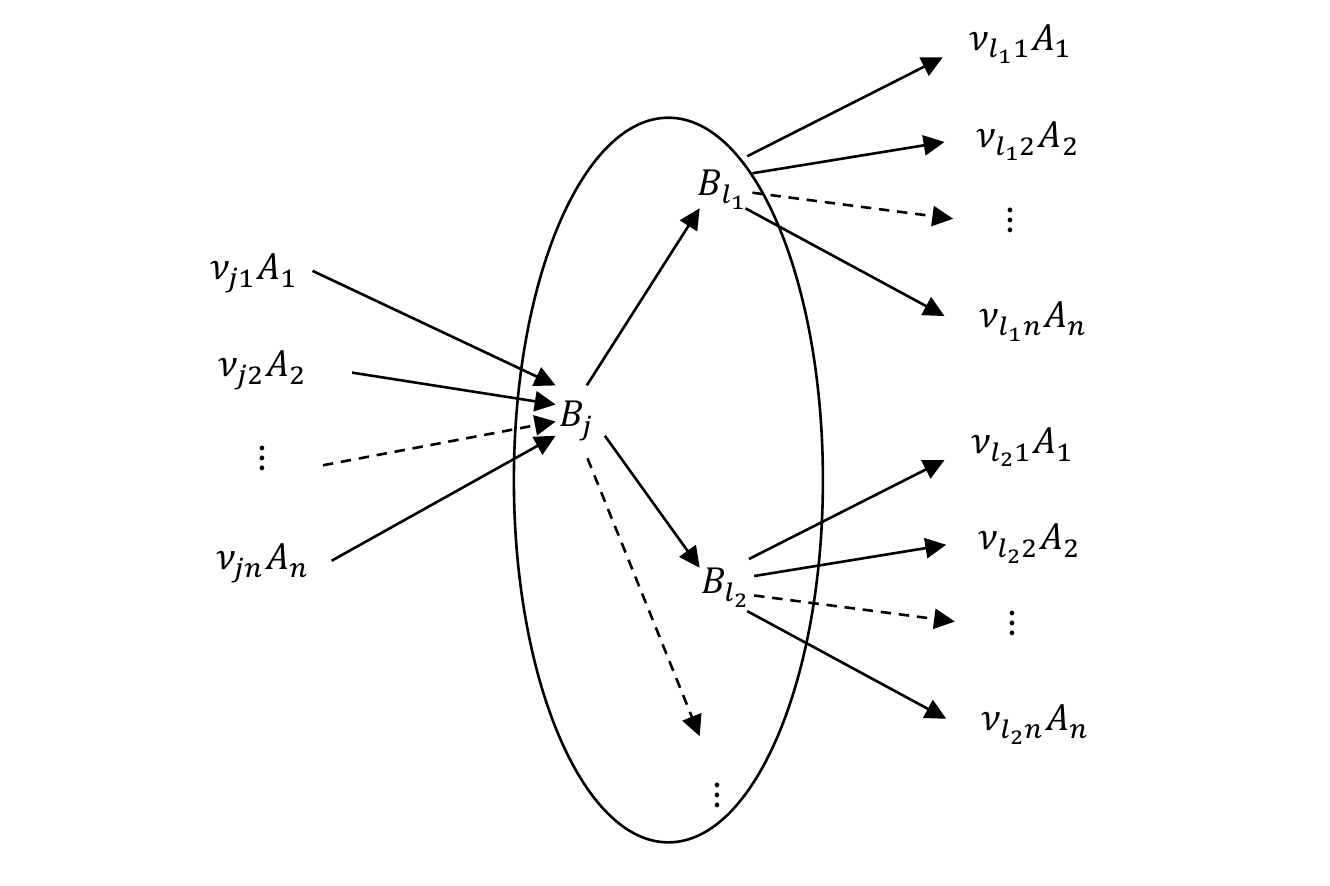}
 \caption{\label{Multichannel}A complex reaction can be decomposed in the groups of transitions associated with compounds $B_j$: $B_j$ is  born and dies  in a reaction (\ref{fastpair}) and can be transformed into several other compounds  $B_{i_k}$. In the MMS asymptotics, the reverse reaction from  (\ref{fastpair}) is much faster than the transitions $B_j \to B_{i_k}$. In our work, we   avoid this assumption. (The reverse transitions $ B_{i_k}  \to B_j$ are also possible but are included in the schemes associated with  $ B_{i_k}$.)}}
\end{figure}

The well founded combination of the generalized MAL with the complex or detailed balance conditions follows from the asymptotic assumptions presented in Fig.~\ref{StueckAsympt}a. These assumptions seem quite realistic, except for one weak point. Fast equilibrium
\begin{equation}\label{fastpair}
\sum_i \nu_{ji}A_i\rightleftharpoons B_j
\end{equation}
assumes that the inverse reaction there is much faster than the transitions between compounds and the lifetime of  $ B_j$ is determined by its decomposition into $\sum_i \nu_{ji}A_i$ (Fig.~\ref{Multichannel}). In enzyme kinetics, such an assumption was abolished by Briggs and Haldane in 1925 \cite{BriggsHaldane1925}.
They found the reaction rate for simple enzyme reaction on the basis of a single assumption: small concentrations of intermediates.  Ironically, the Briggs--Hadane (BH) formula was called the Michaelis--Menten equation. Detailed analysis of the BH  asymptotics was performed in \cite{Segel89}. In the generalized MAL, this analysis is still the open problem.

The MMS approach to exclusion of intermediates is very attractive because it needs no assumption about the rates of the non-linear reactions $\sum_i \nu_{ji}A_i\to B_j$. These non-linear rates are excluded by the fast equilibrium condition and the resulting generalized MAL formulas are based on thermodynamics. The BH approach used the complete kinetic equations as a starting point and serves for the further simplification of {\em existent} MAL equations. But what can we do if the non-linear kinetic law is unknown and there is no fast equilibria between input reagents and intermediate (i.e. the assumption 4 can be wrong)? This is the problem.

\section{Linearisation near partial equilibria}

The first step can be very natural. Let us linearise the kinetic equation for the elementary reaction $\sum_i \nu_{ji}A_i\rightleftharpoons B_j$ near its equilibrium. Because the motion in this reaction is one-dimensional, the linearised equations will have the form of the relaxation time approximation for all non-linear kinetic laws. Therefore, this type of approximation near equilibrium has the same absolute validity as non-equilibrium thermodynamics.

If the concentration $\varsigma_j$ and $ \varsigma^*_j$ are small and we neglect in the minimization the second order quantities   $\varsigma_j \partial \varsigma_j^*(c,T)/\partial c_i$ then the partial  equilibrium concentrations of $B_i$  has a simple form
\begin{equation}\label{equilibrationEq}\boxed{
\varsigma_j^{\rm peq}(c,T)=\varsigma^*_j(c,T)\exp\left(\frac{\sum_i \nu_{ji} \mu_i(c,T)}{RT}\right),}
\end{equation}
where superscript `peq' means partial equilibrium.
The linearized reaction rate for (\ref{fastpair}) is
\begin{equation}\label{reactratelin}
w_j(c,T)=\frac{1}{\tau_j}( \varsigma_j^{\rm peq}(c,T)-\varsigma_j) .
\end{equation}

\begin{equation}\label{newkin}
\begin{split}
\frac{dc_i}{dt}&=-\sum_j \nu_{ji} w_j,\\
\frac{d\varsigma_j}{dt}&=w_j +\sum_{l,\, l\neq j} (\kappa_{jl}\varsigma_l - \kappa_{lj}\varsigma_j).
\end{split}
\end{equation}

The $H$-theorem for these equations (\ref{newkin}) follows from the general results \cite{Gorban2019}: If (i) the  density of reagents' free energy $f_A(c,T)$ (see (\ref{FreeEn2})) is a convex function of $c$, (ii) the partial equilibria in (\ref{reactratelin}) are defined by conditional minimization of  $f(c,\varsigma,T)$  (\ref{FreeEn2}), and (iii) the standard equilibrium $\varsigma^*$ is an equilibrium point of the Markovian kinetics of compounds, then the free energy density $f(c,\varsigma,T)$ does not increase in time on solutions of  (\ref{newkin}).

\section{Small parameter and quasi steady state asymptotics}

Let us introduce explicitly a formal small parameter $\varepsilon$ in (\ref{equilibrationEq}), (\ref{reactratelin}), and (\ref{newkin}):

\begin{equation}\label{equilibrationEqVE}
\varsigma_j^{\rm peq} =\varepsilon \varsigma^*_j \exp\left(\frac{\sum_i \nu_{ji} \mu_i }{RT}\right);
\end{equation}

\begin{equation}\label{reactratelinVE}
w_j =\frac{1}{\varepsilon\tau_j}\left( \varsigma_j^{\rm peq}-\varsigma_j\right) .
\end{equation}

\begin{equation}\label{newkinVE}
\begin{split}
\frac{dc_i}{dt}=&-\sum_j \nu_{ji}\frac{1}{\varepsilon\tau_j}\left( \varsigma_j^{\rm peq}-\varsigma_j\right) ,\\
\frac{d\varsigma_j}{dt}=&\frac{1}{\varepsilon\tau_j}\left( \varsigma_j^{\rm peq}-\varsigma_j\right) + \frac{1}{\varepsilon}\sum_{l, \, l\neq j} (\kappa_{jl}\varsigma_l - \kappa_{lj}\varsigma_j).
\end{split}
\end{equation}

For explicit separation of slow and fast variables a simple linear transformation of concentrations $c_i$ is needed \cite{GorbanShahzad2011}. This transformation excludes reaction rates $w_j$ from the slow equations. The new variables are
$b_i=c_i+\sum_j \nu_{ji}  \varsigma_j$. For them, the equations (\ref{newkin}) have the form

\begin{equation}\label{newkinV}
\begin{split}
\frac{db_i}{dt}&=\sum_{jl, \, j\neq l}  (\kappa_{jl}\varsigma_l - \kappa_{lj}\varsigma_j)\nu_{ji} ,\\
\frac{d\varsigma_j}{dt}&=w_j +\sum_{l, \, l\neq j} (\kappa_{jl}\varsigma_l - \kappa_{lj}\varsigma_j).
\end{split}
\end{equation}
With explicit  small parameters these equations are

\begin{equation}\label{newkinVVE}
\begin{split}
\frac{db_i}{dt}=&\frac{1}{\varepsilon}\sum_{jl, \, j\neq l}  (\kappa_{jl}\varsigma_l - \kappa_{lj}\varsigma_j)\nu_{ji} ,\\
\frac{d\varsigma_j}{dt}=&\frac{1}{\varepsilon\tau_j}\left( \varsigma^{\rm peq}_j-\varsigma_j\right)+\frac{1}{\varepsilon}\sum_{l,\, l\neq j} (\kappa_{jl}\varsigma_l - \kappa_{lj}\varsigma_j).
\end{split}
\end{equation}
Here,  the relaxation times for the equilibration reactions are $\varepsilon\tau_j$ and $\tau_j$  do not depend on $\varepsilon$.

{ 
It is important to stress that the linearization of the equations for compounds in (\ref{newkinV}), (\ref{newkinVVE}) is performed near zero concentration. Any physical kinetics of transitions between compounds should satisfy two conditions: (i) preservation of material balance and (ii) preservation of positivity of concentrations. Therefore, linearization at zero  satisfies the same conditions.  Indeed, if a solution of linearized equation with positive initial state violates one of these conditions, then the solution of the original nonlinear equations with the same initial state also violates the same condition, may be, after rescaling of the initial state by multiplying by a sufficiently small parameter.   If  linearization violates positivity or balance then  higher order terms cannot correct it near zero. A linear system that preserves positivity and balance is a Markov chain (first order kinetics).}

Let  $K$ be a matrix of kinetic coefficients for transitions between compounds:
$K_{jl}=\kappa_{jl}$ for $j\neq l$ and $K_{jj}=-\sum_i  \kappa_{ji}$. This matrix has non-negative non-diagonal elements,  non-positive diagonal and the sum of the elements in columns is zero. Therefore, according to Gershgorin's circle theorem, all the eigenvalues have  non-positive real parts. Let ${\rm diag}[d_i]$ be the diagonal matrix with diagonal elements $d_i$. The matrix  $K-{\rm diag}[{1}/{\tau_i}]$  has non-negative non-diagonal elements,  non-positive diagonal and the sum of the elements in each column is strictly negative. Therefore, the eigenvalues of this matrix have negative real parts (Gershgorin's theorem). In the vector form, the second equation in (\ref{newkinVVE}) is
$$\frac{d\varsigma }{dt}=\frac{1}{\varepsilon}{\rm diag}\left[\frac{1}{\tau_i}\right] \varsigma^{\rm peq}
- \frac{1}{\varepsilon}\left({\rm diag}\left[\frac{1}{\tau_i}\right]-K\right)\varsigma$$

For given $\mu_i$, $T$, the subsystem for concentration of compounds quickly converges to the quasi steady state (qss). (For detailed description of qss and pss approximations, their similarities and differences we refer to review \cite{GorbanCurrOp2018}.)
$$\varsigma^{\rm qss}(\mu, T)=\left({\rm diag}\left[\frac{1}{\tau_i}\right]- K\right)^{-1}{\rm diag}\left[\frac{1}{\tau_i}\right]\varsigma^{\rm peq}(\mu, T)$$
that is $\varepsilon$-small because $\varsigma^{\rm eq}$ (\ref{equilibrationEqVE}) is $\varepsilon$-small. Let us notice a simplification:
\begin{equation}\label{entanglementMatrix}
\left({\rm diag}\left[\frac{1}{\tau_i}\right]- K\right)^{-1}{\rm diag}\left[\frac{1}{\tau_i}\right]=\left(1- {\rm diag}[\tau_i] K\right)^{-1}.
\end{equation}
If $\varsigma$ is $\varepsilon$-close to $\varsigma^{\rm qss}(\mu, T)$ then it is $\varepsilon$-small and the right-hand side of equations for $b_i$ in (\ref{newkinVVE}) is not fast (in this case, it has zero or even higher order in  $\varepsilon$). Thus, the Tikhonov theorem can be applied and the qss approximation is valid for small  $\varepsilon$. We can take into account that the difference $b_i - c_i$ is  $\varepsilon$-small and write the resulting kinetic equation that are asymptotically valid for  $\varepsilon \to 0$:
\begin{equation}\label{result1}
\begin{split}
&\frac{dc_i}{dt}= \frac{1}{\varepsilon}\sum_{jl, \, j\neq l} \left(\kappa_{jl}\varsigma^{\rm qss}_l (\mu, T)- \kappa_{lj}\varsigma^{\rm qss}_j(\mu, T)\right)  \nu_{ji} ;\\
&\varsigma^{\rm qss}(\mu, T) =\left(1- {\rm diag}[\tau_i] K\right)^{-1}\varsigma^{\rm peq}(\mu, T).
\end{split}
\end{equation}

Equations for  $c$ can be modified by changing the indexes in double summation:

\begin{equation}\label{result2}\boxed{
\begin{split}
&\frac{dc_i}{dt}= \frac{1}{\varepsilon}\sum_{jl, \, j\neq l}  \kappa_{jl}\varsigma^{\rm qss}_l (\mu, T) (\nu_{ji}-\nu_{li}) ;\\
&\varsigma^{\rm qss}(\mu, T) =\left(1- {\rm diag}[\tau_i] K\right)^{-1}\varsigma^{\rm peq}(\mu, T).
\end{split}}
\end{equation}

Each term in the right hand side of equations for $c$ in (\ref{result2}) can be represented in the standard form of the sums over reactions. For each pair $j,l$ consider elementary reaction (\ref{elementary reaction}) with $\alpha_{\rho i}=\nu_{l i}$ and $\beta_{\rho i}=\nu_{ji}$. Then, according to (\ref{result2}), equation for $c$ is chemical kinetic equation with reaction rates
\begin{equation}\label{qssreactionrate}
r_{\rho}=\frac{1}{\varepsilon} \kappa_{jl}\varsigma^{\rm qss}_l (\mu, T).
\end{equation}

The qss concentrations of compounds are linear combinations of the peq concentrations. Therefore, the reaction rate  for each reaction (\ref{qssreactionrate}) is combined from the same terms as the generalized MAL reaction rates. The matrix of coefficients of these combinations is

$$E=\left(1- {\rm diag}[\tau_i] K\right)^{-1}.$$
Let us call it the {\em entanglement matrix}.

\section{The MMS limit and the classical generalized MAL}

To obtain the MMS asymptotic  formulas, we may assume that $\tau_i$ are also small. That is, there exists an additional small parameter $\delta>0$ and the relaxation times for the equilibration $\sum_i \nu_{ji}A_i\rightleftharpoons B_j$ is $\delta \varepsilon \tau_j$.

The MMS asymptotics with $\delta \to 0$ gives the generalized MAL reaction rates in the assumption of partial equilibria between compounds $B_j$ and the corresponding complexes $\sum_i \nu_{ji}A_i$:
\begin{equation}\label{peqreactionrate}
\begin{split}
r_{\rho}^{\rm peq}=&\frac{1}{\varepsilon} \kappa_{jl}\varsigma^{\rm peq}_l(\mu, T)= \\
&\frac{1}{\varepsilon} \kappa_{jl}\varsigma^*_l(c,T)\exp\left(\frac{\sum_i \nu_{ji} \mu_i(c,T)}{RT}\right).
\end{split}
\end{equation}
If we use the notation $\varphi_{\rho}=\frac{1}{\varepsilon} \kappa_{jl}\varsigma^*_j(c,T)$ then we obtain the standard generalized MAL factorization of reaction rate onto kinetic and thermodynamic factors (\ref{GMAL}). The semidetailed balance condition (\ref{complexbalance}) is a straightforward consequence of the assumption that the standard thermodynamic equilibrium  $\varsigma^*$ is also the equilibrium of the Markov kinetics of compounds: for all $j$
\begin{equation}\label{MarkovBalance}
\sum_l(\kappa_{jl}\varsigma_l^* - \kappa_{lj}\varsigma_j^*)=0.
\end{equation}

This balance condition means that the microscopic parameters $\kappa_{jl}$ and $\varsigma^*$ are not independent and the constraints (\ref{MarkovBalance}) are nonlinear (bilinear). Happily, in the macroscopic expressions for generalized MAL, like (\ref{peqreactionrate}), these parameters participate together, in a product $\xi_{jl}= \kappa_{jl}\varsigma^*_l$. Parameters $\xi_{jl}$ ($\xi_{jl}\geq 0$, $j\neq l$) have much simpler linear a priory constraints:  for all $j$
\begin{equation}\label{MarkovBalance2}
\sum_l (\xi_{jl} - \xi_{lj})=0.
\end{equation}

 For the generalized MAL, the values of $\xi_{jl}$ can be extracted from the macroscopic reaction rates $r_{\rho}$ (\ref{peqreactionrate}) and the thermodynamic data about reagents $A_i$.

It is impossible to extract the all the microscopic data about $\kappa_{jl}$ and $\varsigma$ from the observation of the macroscopic reaction rates $r_{\rho}$. Nevertheless, the ratio of some microscopic constants can be found. If two elementary reactions have the same input complex with the coefficients $\nu_{li}$ and the output complexes have the  coefficients $\nu_{ji}$ and $n_{qi}$  the ratio of the corresponding reaction rates coincides with  the ratio of the compounds reaction rate constants $\kappa_{jl}/\kappa_{ql}$. According to (\ref{qssreactionrate}) and (\ref{peqreactionrate}), this is true both for the peq and qss reaction rates.

\section{A simple example of entangled MAL}

The structure of the entanglement matrix is closely connected to the representation of the macroscopic reaction mechanism as transformation of complexes. Let us introduce this formalism \cite{HornJackson1972} with the corresponding notations.  Each formal sum in the elementary reactions (\ref{elementary reaction}) is called a complex: $\Theta_l=\sum_i \nu_{li}A_i$. The same complex can participate in several reactions. A complex reaction can be represented as a oriented graph of transition between complexes.

Let us consider a simple example following \cite{GorbanShahzad2011,Conaireatal2004}: 18 elementary reactions (9 pairs
of mutually reverse reactions) from the hydrogen combustion mechanism.
\begin{equation*}
\begin{array}{ll}
 {\rm H + O_2 \rightleftharpoons O + OH;   }& {\rm O + H_2
\rightleftharpoons H + OH; }\\
  {\rm  OH + H_2 \rightleftharpoons H + H_2O;}  &{\rm   O + H_2O \rightleftharpoons 2OH;}\\
 {\rm  HO_2 + H \rightleftharpoons H_2 + O_2;}   &{\rm  HO_2 + H \rightleftharpoons 2OH;}\\
 {\rm  H + OH +M \rightleftharpoons H_2O +M;} & {\rm  H + O_2 +M \rightleftharpoons HO_2 +M; }\\
  {\rm H_2O_2 + H \rightleftharpoons H_2 + HO_2}&
\end{array}
\end{equation*}
There are 16 different complexes here:
\begin{eqnarray*}
&&{\rm \Theta_1=H + O_2, \, \Theta_2= O + OH, } \\
&&{\rm \Theta_3=O + H_2, \, \Theta_4=H+OH, } \\
&&{\rm \Theta_5 =OH + H_2, \, \Theta_6=H + H_2O,} \\
&&{\rm\Theta_7= O + H_2O, \, \Theta_8=2OH,} \\
&&{\rm \Theta_9=HO_2 + H, \, \Theta_{10}=H_2 + O_2, }\\
&&{\rm \Theta_{11}=H + OH +M, \, \Theta_{12}=H_2O +M,}\\
&&{\rm \Theta_{13}=H + O_2 +M,\,  \Theta_{14}=HO_2 +M, }\\
&&{\rm \Theta_{15}=H_2O_2 + H, \,  \Theta_{16}=H_2 + HO_2}.
\end{eqnarray*}
The reaction mechanism can be represented as
\begin{eqnarray*}
&& \Theta_1 \rightleftharpoons \Theta_2, \,  \Theta_3 \rightleftharpoons \Theta_4, \, \Theta_5\rightleftharpoons
\Theta_6,\\
&& \Theta_7\rightleftharpoons \Theta_8
\rightleftharpoons \Theta_9 \leftrightharpoons \Theta_{10},  \\
&&  \Theta_{11}\rightleftharpoons \Theta_{12},
\,\Theta_{13}\rightleftharpoons \Theta_{14}, \,
\Theta_{15}\rightleftharpoons \Theta_{16}\,
\end{eqnarray*}
This graph of transformation of complexes has
a very simple structure: There are five isolated pairs of
complexes and one connected group of four complexes.

Let us rewrite the basic formulas in these notations. For each reaction $ \Theta_l \to  \Theta_j$ consider the generalized MAL reaction rate (\ref{GMAL})
\begin{equation}\label{GMAL1}
r_{jl}=\varphi_{jl}  \exp\left(\frac{\sum_i \nu_{ji} \mu_i(c,T)}{RT}\right).
\end{equation}
(The standard order of indexes is used, $r_{j\leftarrow l}$.) For the perfect systems the free energy density is
$$f(c,T)=RT\sum_i c_i \left(\ln\left(\frac{c_i}{c_i^*}\right)-1\right),$$
where $c_i^*$ is a standard equilibrium (an equilibrium point for a priori selected values of linear balances).

The Bolltzmann factor for the perfect system is:
$$\exp\left(\frac{\sum_i \nu_{ji} \mu_i(c,T)}{RT}\right)=\prod_i \left(\frac{c_i}{c_i^*}\right)^{\nu_{ji}}$$
under the standard assumption that $c^{\alpha}=1$ if $\alpha=0$ and $c\geq 0$. Then the generalized MAL turns into the classical MAL
$$r_{jl}=\varphi_{jl}\prod_i \left(\frac{c_i}{c_i^*}\right)^{\nu_{ji}}$$
Under assumption (\ref{complexbalance}):  for all $j$
$$\sum_j \varphi_{jl}\equiv \sum_l \varphi_{lj},$$
or, with the systems with microreversibility (the most common case in physics and chemistry \cite{Gorban2014}):  for pairs all $j, l$ ($j\neq l$)
$$\varphi_{jl}\equiv \varphi_{lj}.$$
The non-negative quantities $\varphi_{jl}$ are not compulsory constant. Their  detailed dependence on the concentrations, temperature, or other conditions is beyond the scope of this work.

For the entanglement example, we select a system with five components $A_{1-5}$, three complexes,  $\Theta_1 =A_1+A_2$, $\Theta_2 =2A_3$, $\Theta_3 =A_4+A_5$, and two reversible reactions:
$$ \Theta_1 \rightleftharpoons \Theta_2 \rightleftharpoons \Theta_3.$$
We can find such fragments in many complex reactions, for example, in the hydrogen combustion mechanism mentioned above.

According to the transient state modeling (Figs.~\ref{StueckAsympt}, \ref{Multichannel}), there is one compound for each complex, $B_i$ ($i=1,2,3$). The complete scheme of reactions with compounds consists of five reversible reactions:  $ \Theta_i \rightleftharpoons  B_i$ ($i=1,2,3$) and $ B_1 \rightleftharpoons B_2 \rightleftharpoons B_3$.

For kinetic model we need the following data: five standard equilibrium concentrations for reagents, $c_{1-5}^*$, three standard equilibrium concentrations for compounds, $\varsigma_i^*$, four reaction rate constants for transitions between compounds, $\kappa_{21}$, $\kappa_{12}$, $\kappa_{32}$, and $\kappa_{23}$, and three relaxation times for the equilibraton between the complexes and compounds, $\tau_{1-3}$ (\ref{reactratelin}).  These data are not independent and the balance conditions should hold (\ref{MarkovBalance}). We assume even stronger physical conditions, detailed balance:
\begin{equation}\label{detbalEx}
\kappa_{21}\varsigma_1^*= \kappa_{12}\varsigma_2^*(=w^*_1) \; \mbox{ and } \; \kappa_{32}\varsigma_2^*= \kappa_{23}\varsigma_3^*(=w^*_2),
\end{equation}
where $w^*_{1,2}$ are the equilibrium fluxes between the corresponding compounds.

The partial equilibrium concentrations of compounds are (\ref{equilibrationEq}):
\begin{equation}\label{equilibrationEqEx}
\varsigma_j^{\rm peq} =\varsigma^*_j \prod_i \left(\frac{c_i}{c_i^*}\right)^{\nu_{ji}},
\end{equation}
 in particular,
 $$\varsigma_1^{\rm peq}=\varsigma^*_1\frac{c_1}{c_1^*}\frac{c_2}{c_2^*}, \; \varsigma_2^{\rm peq}=\varsigma^*_2\left(\frac{c_3}{c_3^*}\right)^2, \;
 \varsigma_3^{\rm peq}=\varsigma^*_3\frac{c_4}{c_4^*}\frac{c_5}{c_5^*}. $$
 For the generalized MAL reaction rates, these expressions together with the detailed balance conditions (\ref{detbalEx}) give (\ref{peqreactionrate}):
 \begin{equation}\label{peqreactratesEx}
 \begin{split}
 r_{21}^{\rm peq}&=w_1^*\frac{c_1}{c_1^*}\frac{c_2}{c_2^*}, \; r_{12}^{\rm peq}=w^*_1 \left(\frac{c_3}{c_3^*}\right)^2, \\
 r_{32}^{\rm peq}&=w^*_2\left(\frac{c_3}{c_3^*}\right)^2, \; r_{23}^{\rm peq}=w^*_2 \frac{c_4}{c_4^*}\frac{c_5}{c_5^*}.
 \end{split}
 \end{equation}
 The entanglement matrix is  $E=\left(1- {\rm diag}[\tau_i] K\right)^{-1}$:
 \begin{equation*}\label{entanglementEx}
 E= \left[
\begin{array}{ccc}
 1+\tau_1\kappa_{21} & - \tau_1 \kappa_{12} & 0 \\
 -\tau_2 \kappa_{21} & 1+\tau_2(\kappa_{12}+\kappa_{32}) & -\tau_2 \kappa_{23} \\
 0                           & -\tau_3 \kappa_{32} & 1+ \tau_3 \kappa_{23}
 \end{array}
 \right]^{-1}    \, .
 \end{equation*}
 The vector of qss concentrations of compounds is $\varsigma ^{\rm qss}=E\varsigma^{\rm peq}$. The qss reaction rates are: $r_{jl}^{\rm qss}=\kappa_{jl}\varsigma ^{\rm qss}_l$. It is straightforward to get explicit formulas for reaction rates, but they are quite cumbersome, so we will limit ourselves to a  numerical example. Let $c_{1_5}^*=1$,  $\varsigma^*_{1-3}=1$, $\kappa_{ij}=1$, and $\tau_{1-3}=1$ (here we omit the small parameter, since it is vanished in microscopic quantities). For these data, $w_1^*=w_2^*=w_3^*=1$,
  $$\varsigma_1^{\rm peq}={c_1}{c_2}, \; \varsigma_2^{\rm peq}={c_3}^2, \;  \varsigma_3^{\rm peq}={c_4}{c_5}; $$

   \begin{equation*}
 E=\left[
\begin{array}{ccc}
2 & -1 & 0\\
 -1 & 3 & -1\\
0   &-1 & 2
 \end{array}
 \right]^{-1}=
 \frac{1}{8} \left[
\begin{array}{ccc}
5 & 2 & 1\\
 2 & 4 & 2\\
1   &2 & 5
 \end{array}
 \right]     \, ;
 \end{equation*}

     \begin{equation*}
\varsigma^{\rm qss}=\frac{1}{8} \left[
\begin{array}{c }
5{c_1}{c_2}+2{c_3}^2+ {c_4}{c_5}\\
 2{c_1}{c_2}+ 4{c_3}^2 + 2 {c_4}{c_5}\\
{c_1}{c_2} +2{c_3}^2+ 5{c_4}{c_5}
 \end{array}
 \right]     \, ;
 \end{equation*}

  \begin{equation*}
  \begin{split}
 & r_{21}^{\rm qss}=\frac{1}{8}(5{c_1}{c_2}+2{c_3}^2+ {c_4}{c_5}), \;  r_{12}^{\rm qss}= \frac{1}{8}(2{c_1}{c_2}+ 4{c_3}^2 + 2 {c_4}{c_5}),\; \\
 &  r_{32}^{\rm qss} =\frac{1}{8}(2{c_1}{c_2}+ 4{c_3}^2 + 2 {c_4}{c_5}),\;  r_{23}^{\rm qss}= \frac{1}{8}({c_1}{c_2} +2{c_3}^2+ 5{c_4}{c_5}).
 \end{split}
  \end{equation*}

   \begin{equation*}
  \begin{split}
  \frac{d c_1}{dt}=&\frac{d c_2}{dt}=  \frac{1}{8}(-3{c_1}{c_2}+2{c_3}^2+ {c_4}{c_5}), \\
  \frac{d c_3}{dt}=&  \frac{1}{4}(2{c_1}{c_2}-4{c_3}^2+2{c_4}{c_5})  \\
   \frac{d c_4}{dt}=&\frac{d c_5}{dt}=   \frac{1}{8}({c_1}{c_2}+ 2{c_3}^2 -3 {c_4}{c_5}).
 \end{split}
  \end{equation*}

  If we decipher this system according to the standard (textbook \cite{MarinYab2019}) MAL kinetics then we find  three reversible elementary reactions
  $$ \Theta_1 \rightleftharpoons \Theta_2 \rightleftharpoons \Theta_3 \rightleftharpoons \Theta_1 .$$
Initial two reversible elementary reactions turned into three ones. The reaction rate constants also changed. In the stoichiometric form, these reactions with the new reaction rates are listed below
\begin{enumerate}
 \item $A_1+A_2  \rightleftharpoons 2A_3$, $k_1^+=\frac{1}{4}$, $k_1^-=\frac{1}{4}$;
 \item $2A_3  \rightleftharpoons A_4+A_5$, $k_2^+=\frac{1}{4}$, $k_2^-=\frac{1}{4}$;
  \item $A_4+A_5  \rightleftharpoons A_1+A_2 $, $k_2^+=\frac{1}{8}$, $k_2^-=\frac{1}{8}$.
\end{enumerate}

We see that the connected component of the complex transition graph turned into the complete digraph. This is the universal effect of the entanglement matrix.  The equilibrium did not change.

\section{Entanglement matrix in the first approximation}

The MMS asymptotics (Fig.~\ref{StueckAsympt}a) assumes that equilibration of fast equilibria is infinitely much faster than other transitions between compounds (the products $\tau_k\kappa_{ji}\to 0$ for all $i,j,k$).  We rejected this assumption (Fig.~\ref{StueckAsympt}b) and, in combination with linearisation  near partial equilibria, obtained the entanglement matrix (\ref{entanglementMatrix}), (\ref{result2}). Let us analyze an intermediate assumption that   equilibration of fast equilibria is  faster than other transitions between compounds but not infinitely faster. Introduce a formal small parameter $\delta$.
$$E=\left(1- \delta{\rm diag}[\tau_i] K\right)^{-1}.$$
For sufficiently small $\delta$ the following series for $E$ converges:
$$E=1+\delta{\rm diag}[\tau_i] K+\delta^2({\rm diag}[\tau_i] K)^2+\ldots .$$
In the first approximation,
$$E=1+\delta{\rm diag}[\tau_i] K +o(\delta).$$
If we introduce such small parameter into the simple example from the previous section then we obtain:
   \begin{equation*}
 E=1+ \delta \left[
\begin{array}{ccc}
1 & -1 & 0\\
 -1 & 2 & -1\\
0   &-1 & 1
 \end{array}
 \right] +o(\delta)\, ;
 \end{equation*}

       \begin{equation*}
\varsigma^{\rm qss}= \left[
\begin{array}{c }
(1+\delta){c_1}{c_2}-\delta {c_3}^2,\\
 -\delta {c_1}{c_2} + (1+2\delta){c_3}^2 -\delta{c_4}{c_5})\\
-\delta{c_3}^2 + (1+\delta){c_4}{c_5}
 \end{array}
 \right]     \, ;
 \end{equation*}
 \begin{equation*}
 \begin{split}
 & r_{21}^{\rm qss}=(1+\delta){c_1}{c_2}-\delta {c_3}^2, \;  r_{12}^{\rm qss}= -\delta {c_1}{c_2} + (1+2\delta){c_3}^2 -\delta{c_4}{c_5}),\; \\
 &  r_{32}^{\rm qss} =-\delta {c_1}{c_2} + (1+2\delta){c_3}^2 -\delta{c_4}{c_5}),\;  r_{23}^{\rm qss}= -\delta{c_3}^2 + (1+\delta){c_4}{c_5}.
\end{split}
 \end{equation*}
Again, the reaction rates become linear combinations of known MAL dependencies. The approximate formulas for small $\delta$ can lead to the positivity loss when some of macroscopic concentrations $c_i$ become $\delta$-small.

\section{Discussion, conclusion and outlook}

The century-old  classical MMS asymptotic approach to  derivation of (generalized) Mass Action Law (MAL) implies  two properties of intermediate transition states:
\begin{description}
\item[Quasi steady state.] The intermediate  states  have short life time and are present in much smaller concentrations than the main reagents.
\item[Fast equilibrium.] The intermediate  states are in  partial equilibrium with the initial reagents of the elementary reaction. This means that  the reverse decomposition of the intermediates is much faster than its transition through other channels to the products.
\end{description}
The transition state (activation complex) theory also relies on these assumptions. The intermediate  states are named differently by different groups of researchers: the transition states, the activation complexes or just compounds. We  called them `compounds' employing the Michaelis--Menten terminology.

The  classical approaches have  an important advantage: there is no need to postulate a kinetic law for the birth of the transition state (compounds) from the reagents in non-linear reactions. This non-linear kinetic law follows from the thermodynamic description of the partial equilibrium and then the  transformations of transition states is described by linear kinetics (Markov chains). The kinetic constants for this linear kinetics can be evaluated (the  transition state or activation complex theory) or extracted from combination of experimental data and theoretic estimates. The structure of the kinetic law will be the same: the generalized MAL.

It is possible to relax the assumption of fast equilibrium without postulating a law of non-linear reactions. For this purpose, we can just linearize the {\em still unknown} equation of the birth of compounds from combinations of initial reagents. The single elementary reaction goes along one stoichiometric vector, therefore, the linearized kinetics can be described by one constant -- the relaxation time. Of course this relaxation time may be different for different compounds and also depend  on the conditions.

We demonstrate how substituting of the partial equilibrium assumption by the weaker assumption of the linear kinetics near partial equilibria modifies the reaction rates. In this approach, the asymptotic quasi steady state expressions for reaction rates were produced.   The final reaction rates are combinations of the generalized MAL expressions but for different reactions. We called this effect the `entanglement' of MAL. From the transition state theory point of view, the entanglement matrix $E$ can be considered as the {\em generalized matrix transmission coefficient}.

In particular, any connected component of the digraph of the transitions between complexes is transformed into a complete digraph of reactions. Thus, the set of reactions  $ \rm O + H_2O\rightleftharpoons 2OH \rightleftharpoons  HO_2 + H \leftrightharpoons H_2 + O_2$ should be compulsory supplemented to the complete digraph by three reversible reactions
$ \rm O + H_2O \rightleftharpoons   HO_2 + H$, $ \rm O + H_2O \leftrightharpoons H_2 + O_2$, and $ \rm 2OH \rightleftharpoons H_2 + O_2$.

The approach based on the quasi steady state assumption that the intermediates  have short life time and are present in small concentrations works well if the microkinetic equations for birth of eacch compound are linearized near the corresponding partial equilibria. In this approximation, the kinetic law is the `entangled MAL' combined from the same terms. This approach can be incorporated in modern advances framework  of multiscale non-equilibrium thermodynamics \cite{Grmela2018}. The next step beyond this approximation requires the analysis of microscopic models of elementary reaction phenomena, depends on the details of the interaction of particles, and goes beyond the scope of universal phenomenological theory.

\begin{figure}
\centering{
\includegraphics[width=\textwidth]{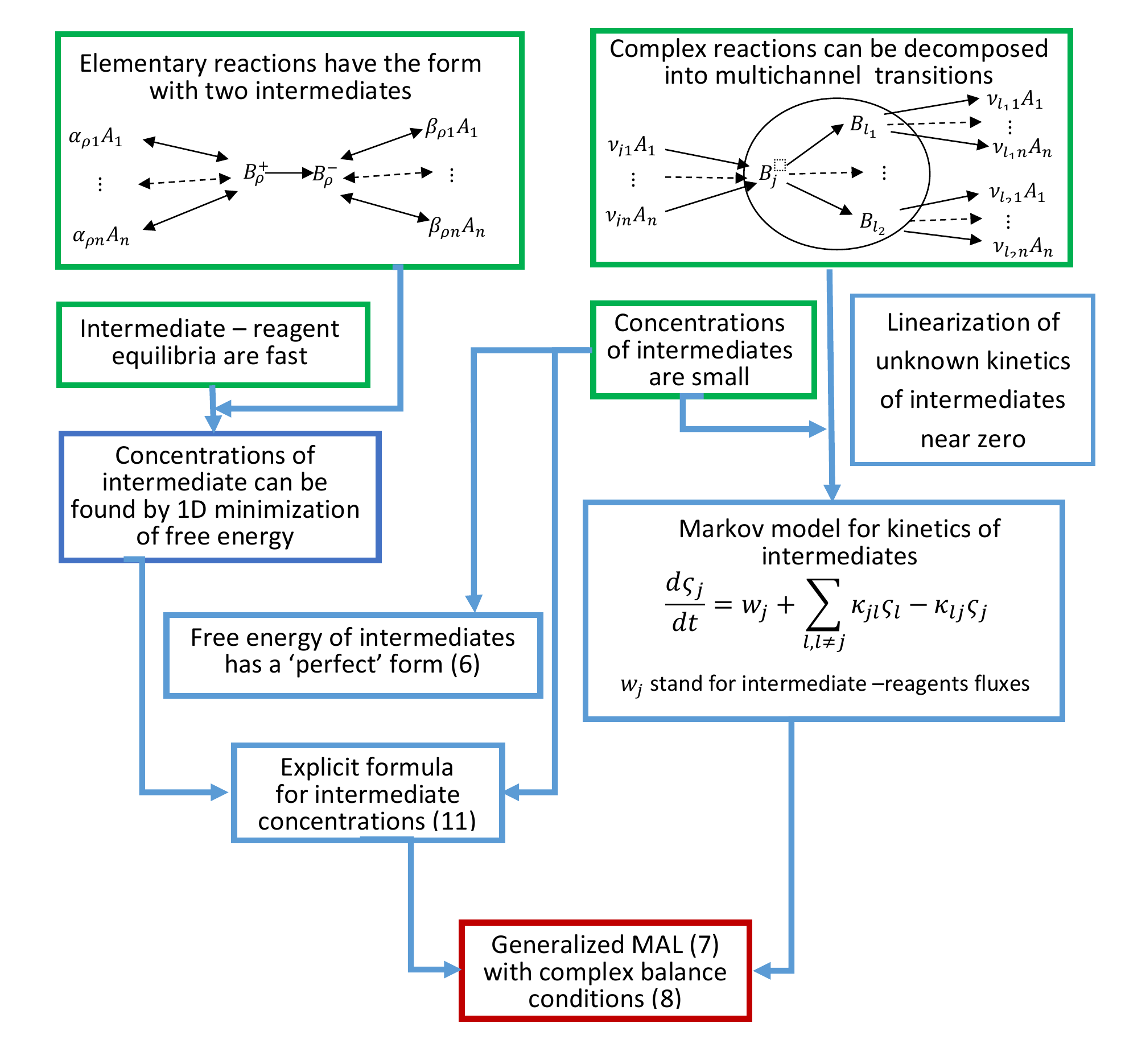}
 \caption{\label{Flowchart1} Main assumptions (green frame), intermediate  statements (blue frames) and the main statement  in Michaelis--Menten--Stueckelberg theorem about generalized MAL with complex balance conditions. The main assumptions are (Fig. \ref{StueckAsympt}a): (i) representation of elementary reactions by schemes with two intermediates, (ii)  decomposition of complex reaction into multichannel transitions, (iii) fast equilibria between reagents and intermediates, and (iv) small  intermediate concentrations. Dashed frame indicates operation of linearization.}}
\end{figure}

\begin{figure}
\centering{
\includegraphics[width=\textwidth]{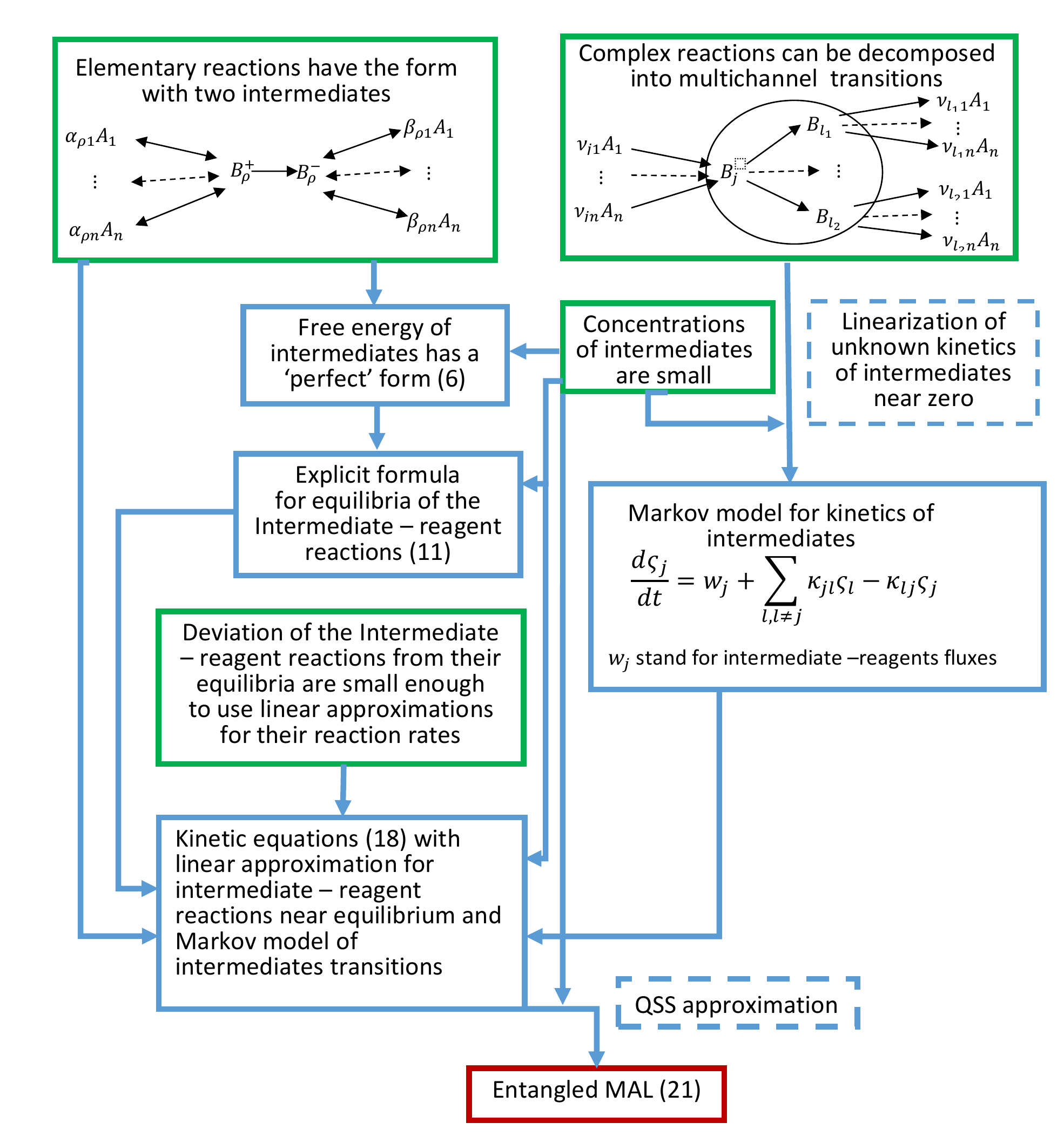}
 \caption{\label{Flowchart2} Main assumptions (green frame) and intermediate  statements (blue frames) that imply generalized entangled MAL  (\ref{result2}) . The main assumptions are (Fig. \ref{StueckAsympt}a): (i) representation of elementary reactions by schemes with two intermediates, (ii)  decomposition of complex reaction into multichannel transitions, (iii) small deviations of fast reactions between reagents and intermediates from their equilibria, and (iv) small  intermediate concentrations. Dashed frame indicates operations of linearization and QSS approximation.}}
\end{figure}

{
 Is there a {\em loophole / logic cycle} in determination of generalized MAL? 
 Michaelis and Menten hypothesized MAL for rapid equilibrium between reagents and compounds that are present in small quantities, and obtained a dynamic MAL for the brutto-reaction. In brief, it can be formulated as a logic cycle: we assumed MAL -- we got MAL, but this is not the case. Michaelis and M assume MAL for the equilibrium between reagents and compounds. This is a consequence of the perfect form of the free energy and does not require any kinetic assumption. Then they used that the intermediate compounds are present in small amounts. This leads to the first order kinetics for their transitions. (Of course, neither Michaelis and Menten nor Stueckelberg performed formal proofs.  For example, they did not prove the correct linear kinetics for a general kinetic system with material balance and positivity preservation near zero concentrations and just wrote the first order kinetics.) The result was the dynamic MAL that described the reaction rate of the overall brutto-reaction, not only the equilibrium. Moreover, Briggs and Haldane demonstrated that if we do not assume that the equilibrium between reagents and intermediate compounds is approaching much faster than the reaction between intermediates, then the MAL for the brutto reaction is not valid even we assume the dynamic MAL for all the intermediate elementary reactions. (The historical paradox -- result of  Briggs and Haldane  is called the Michaelis--Menten formula). Thus, there is no logic cycle here and the assumed MAL for equilibrium did not automatically imply the dynamic MAL for the brutto-reaction. The same is true for the general Michaelis--Menten--Stueckelberg theorem.
 
We represent logical chains of reasoning for obtaining a generalized MAL in the Michaelis-Menten-Stueckelberg theorem and in this work by two  flowcharts (Figs. \ref{Flowchart1}, \ref{Flowchart2}). 

It should be noted that the assumption of small intermediate concentrations is repeatedly used in the work not only directly to calculate the QSS approximation, but also to simplify the free energy function, to explicitly solve the 1D minimization of free energy, which gives us quasi-equilibrium concentrations of intermediate substances, to approximate unknown kinetic equations for compounds near zero concentrations, to derive kinetic equations (\ref{newkinV}), (\ref{newkinVVE}) (here the smallness of the intermediate concentration is used not only to determine the asymptotic limit, but also for the introduction of correct macroscopic variables $b_i$), for some other technical purposes, and finally for the derivation of the QSS asymptotics of kinetic equations, which gives us the main result -- the generalized entangled MAL (\ref{result2}).
Thus, assumption of smallness of concentrations of intermediate compounds is a `hub' on flowcharts  (Figs. \ref{Flowchart1}, \ref{Flowchart2}).
  
 }

There is one more open possibility of universal laws. These are the laws of fluctuations. Two closely related questions appear, up-and down \cite{Ottinger200201,Ottinger200202}:
\begin{description}
\item[Up:] coarse-graining. Given a microscopic model and reasonable set of macroscopic or mesoscopic variables generate the universal thermodynamic structure of the phenomenological equations
and a meaningful coarse-graining map, how can one find
the thermodynamic structure of the macroscopic, phenomenological equation?
\item[Down:] noise enhancement.  Given a phenomenological equation and its thermodynamic structure, how can we add fluctuations in a universal and correct form?
\end{description}

 We demonstrated a step up, coarse-graining with the exclusion of intermediates from macroscopic equations. The result looks quite universal. A corresponding step down is also needed.
 For this purpose, we should employ the general fluctuation-dissipation theorem, which relates the dissipation in kinetic equations to the fluctuation extension of these equations to the Markov process.
 A case study of a simple chemical reaction  $A \leftrightharpoons B$ was presented in \cite{Ottinger200202}. Such an analysis should be extended to the general case.

Another  dimension  of future work is the learning physics from data. Here the interaction between data science and nonequilibrium thermodynamics is twofold. First, we can combine our knowledge of thermodynamic structures, some theoretical estimates and data, and create reliable and accurate models. A better understanding of the underlying physical structures creates more constraints on models and allows for more efficient use of data. As for complicated and sometimes ambiguous calculations of the transition state theory, we can  quote the famous talk of Geoffrey F. Chew  at Lawrence Radiation Laboratory, Berkeley about $S$-matrix theory \cite{Chew1961}: ``a long theoretical chain of calculations can be reinforced at its weak links by a direct infusion of experimental information.''  On another hand,  the the machine learning process  can be considered as a dissipative time evolution. This is the gate for thermodynamics to enter  machine learning. Exclusion of intermediates and other thermodynamic model reduction processes can be considered as a model of learning. The approach to learning via reducing dynamics and associated thermodynamics was proposed recently and illustrated by examples of learning physics from data \cite{Chinesta2019}.

Thus, we see two main directions for further research: `noise enhancement' (adding fluctuations) and `learning physics from data' (combine machine learning and a thermodynamic framework  to effectively use experimental data in modeling).

{
In this work, one fundamental question remained unresolved. What is the relationship between the relaxation times of the equilibria `reactants -- intermediates' and the Markov chain of transitions between intermediates? Or, which is about the same question, what are the relative orders of magnitude between the groups of constants $ 1 / \tau_l $ and $ \kappa_ {ij} $ in equations (\ref{reactratelin}), (\ref{newkin}) and further? In the classical asymptotic  assumption  (Fig. \ref{StueckAsympt} a) relaxation time of fast equilibria  (i.e. $ \tau $) is much smaller (faster) than the relaxation time of the transitions between the intermediates.
No fundamental reasons for this assumption is known other than simplicity and the possibility of explicit solutions. Moreover, in the kinetics of enzymatic reactions and in heterogeneous catalysis, this assumption has long been discarded (after Briggs and Haldane \cite{BriggsHaldane1925}). The relations between these relaxation times and reaction rate constants may be different, and the stages of  creating intermediates  are included in the reaction mechanism without any a priori hypothesis of time separation \cite{YBGE1991}.  The situation with transition state theory is more complicated because the (generalized) dynamic MAL for elementary reactions is not assumed. A detailed microscopic theory of elementary reactions has not yet been developed, and, moreover, we cannot expect a universal answer from this theory. An experimental approach to can be proposed for answering this question. The entangled MAL  (\ref{result2}) leads to some inequalities between observable MAL constants of elementary reactions: if there exists a chain of complex transformations $\Theta_1 \rightleftharpoons \Theta_2 \rightleftharpoons \ldots  \rightleftharpoons  \Theta_m$, then the reaction $\Theta_1  \rightleftharpoons  \Theta_m$ should be also observed and its reaction rate constants cannot be arbitrarily small. The lower bounds for these constants can be evaluated. They depend on relations between the sets for constants $ 1 / \tau_l $ and $ \kappa_ {ij} $ for the chain. So, some of these relations can be estimated experimentally.

}

\section{Acknowledgments}
The work was supported by the University of Leicester and the Ministry of Science and Higher Education of Russian Federation  (Project No. 075-15-2020-808).

\vspace{2cm}

\noindent{\bf Alexander N. Gorban}, DrSc, PhD, Prof,  has held a Personal Chair in applied mathematics at the University of Leicester since 2004. He worked for the Russian Academy of Sciences, Siberian Branch, and ETH Zurich, was a visiting professor and scholar at Clay Mathematics Institute (Cambridge, MA), IHES (Bures-sur-Yvette), Courant Institute (New York), and Isaac Newton Institute (Cambridge, UK). His main research interests are machine learning, data mining and model reduction problems, physical, chemical and biological kinetics, and applied dynamics.

\begin{thebibliography}{10}
\bibitem{Laidler1983}K. Laidler, C. King,  Development of transition-state theory,  J. Phys. Chem. 87 (15) (1983), 2657--2664. \url{https://doi.org/10.1021/j100238a002}

\bibitem{MichaelisMenten1913}L. Michaelis, M. Menten,   Die {K}inetik  der {I}nvertinwirkung, {Biochem. Z.} { 49} (1913),     333--369. [English translation: Biochemistry 50 (2011), 8264--8269  \url{https://doi.org/10.1021/bi201284u}]


 \bibitem{Stueckelberg1952}E.C.G. Stueckelberg,  Theoreme $H$ et
    unitarite de $S$, {Helv. Phys. Acta} {25} (1952), 577--580. \url{http://doi.org/10.5169/seals-112324}

\bibitem{Watanabe1955}S. Watanabe,  Symmetry of physical laws Part I. Symmetry in space-time and balance theorems. Rev. Mod. Phys. 27(1) (1955), 26--39.
\url{https://doi.org/10.1103/RevModPhys.27.26}

\bibitem{Boltzmann1887}L. Boltzmann, Neuer Beweis zweier S\"atze über das W\"armegleichgewicht unter mehratomigen Gasmolek\"ulen. Sitzungsberichte der Kaiserlichen Akademie der Wissenschaften in Wien  95 (2)  (1887), 153--164.

 \bibitem{HornJackson1972}F. Horn, R. Jackson, General mass
    action kinetics. {Arch. Ration. Mech. Anal.} 47 (1972), 81--116. \url{https://doi.org/10.1007/BF00251225}

\bibitem{Feinberg1972}M. Feinberg, Complex balancing in general
    kinetic systems. { Arch. Ration. Mech. Anal.} 49 (1972), 187--194. \url{https://doi.org/10.1007/BF00255665}

\bibitem{GorbanShahzad2011}A.N. Gorban, M. Shahzad, The Michaelis-Menten-Stueckelberg theorem, Entropy  13(5) (2011), 966--1019. \url{https://doi.org/10.3390/e13050966}

\bibitem{GorbanKol2015}A.N. Gorban, V.N. Kolokoltsov, Generalized mass action law and thermodynamics of nonlinear Markov processes.
Math. Model. Nat. Phenom. 10(5) (2015), 16--46. 		\url{https://doi.org/10.1051/mmnp/201510503}

\bibitem{Gorban2014}A.N. Gorban, Detailed balance in micro- and macrokinetics and micro-distinguishability of macro-processes. Results Phys. 4 (2014), 142--147.
\url{https://dx.doi.org/10.1016/j.rinp.2014.09.002}

\bibitem{Perez2017} J.F. Perez-Benito, Some considerations on the fundamentals of chemical kinetics: steady state, quasi-equilibrium, and transition state theory. J. Chem. Educ. 94(9)  (2017), 1238--1246. \url{https://dx.doi.org/10.1021/acs.jchemed.6b00957}

\bibitem{BriggsHaldane1925}G.E. Briggs, J.B.S. Haldane,  A note
    on the kinetics of enzyme action, {Biochem. J.} {19}  (1925),
    338--339. \url{https://doi.org/10.1042/bj0190338}

 \bibitem{Bao2017}J.L. Bao, D.G. Truhlar,   Variational transition state theory: theoretical framework and recent developments. Chem. Soc. Rev. 46(24) (2017), 7548--7596.
\url{https://doi.org/10.1039/c7cs00602k}

  \bibitem{Pollak2018}E. Pollak, Stochastic Transition State Theory. J. Phys. Chem. Lett.  9(20) (2018), 6066--6071
\url{https://doi.org/10.1021/acs.jpclett.8b02712}

\bibitem{Carvalho2017}V.H. Carvalho-Silva, V. Aquilanti,  H.C. de Oliveira,  K.C.  Mundim,  Deformed transition-state theory: Deviation from Arrhenius behavior and application to bimolecular hydrogen transfer reaction rates in the tunneling regime. J. Comput. Chem. 38(3) (2017), 178--188. \url{https://doi.org/10.1002/jcc.24529}

\bibitem{JangVoth2017}S. Jang,  G.A. Voth,   Non-uniqueness of quantum transition state theory and general dividing surfaces in the path integral space. J. Chem. Phys.  146(17) (2017), 174106. \url{https://doi.org/10.1063/1.4982053}

\bibitem{Sharia2016}O. Sharia, G. Henkelman, Analytic dynamical corrections to transition state theory. New J. Phys. 18(1)  (2016), 013023.  \url{https://doi.org/10.1088/1367-2630/18/1/013023}

\bibitem{Gesu2017}G. Di Ges\`{u}, T.  Leli\`{e}vre, D. Le Peutrec, B.  Nectoux,  Jump Markov models and transition state theory: the quasi-stationary distribution approach. Faraday Discuss. 195 (2017), 469--495. \url{https://doi.org/10.1039/C6FD00120C}

\bibitem{GorbanSlow}A.N. Gorban, Singularities of transition processes in dynamical systems: Qualitative theory of critical delays, Electron. J. Diff. Eqns., Monograph 05, 2004. Online: \url{http://ejde.math.unt.edu/Monographs/05/gorban.pdf}

\bibitem{Kurth2017}T. Kittel, J. Heitzig, K. Webster, J. Kurths,  Timing of transients: quantifying reaching times and transient behavior in complex systems.  New J. Phys. 19(8)  (2017), 083005. \url{https://doi.org/10.1088/1367-2630/aa7b61}

\bibitem{GorbanSing2020}A.N. Gorban, Singularities of transient processes in dynamics and beyond, Phys. Life Rev. 32 (2020), 46--49.
\url{https://doi.org/10.1016/j.plrev.2019.12.002}

\bibitem{MarinYab2019}G.B. Marin, G.S. Yablonsky, D. Constales,  Kinetics of chemical reactions: Decoding complexity. John Wiley \& Sons, 2019.

\bibitem{Grmela2018}M. Pavelka, V. Klika, M. Grmela,  Multiscale thermo-dynamics: introduction to GENERIC, De Gruyter, Berlin/Boston, 2018.

\bibitem{Feinberg1972MDD}M. Feinberg,  On chemical kinetics of a certain class.  Arch. Ration. Mech. Anal., 46 (1972), 1--41. \url{https://doi.org/10.1007/BF00251866}

\bibitem{BykGorYab1982MDD}V.I. Bykov, A.N. Gorban, G.S. Yablonskii, Description of nonisothermal reactions in terms of {M}arcelin--de-{D}onder kinetics and its generalizations, React. Kinet. Catal. Lett., 20 (1982), 261--265. \url{https://doi.org/10.1007/BF02066307}

\bibitem{Conaireatal2004}M. \`O Conaire,  H.J. Curran,
    J.M. Simmie, W.J. Pitz,  C.K. Westbrook,
    A comprehensive modeling study of hydrogen oxidation.
    {Int. J. Chem. Kinet.}     {36} (2004),  603--622. \url{https://doi.org/10.1002/kin.20036}

\bibitem{Segel89}L.A. Segel, M. Slemrod,   The
    quasi-steady-state assumption: A case study in
    perturbation. { SIAM Rev.}  {31} (1989), 446--477. \url{https://doi.org/10.1137/1031091}

\bibitem{Gorban2019}A.N. Gorban, Universal Lyapunov functions for non-linear reaction networks. Comm. Nonlinear Sci. Numer. Simulat.  79 (2019),
104910.  \url{https://doi.org/10.1016/j.cnsns.2019.104910}

\bibitem{GorbanCurrOp2018}A.N. Gorban,  Model reduction in chemical dynamics: slow invariant manifolds, singular perturbations, thermodynamic estimates, and analysis of reaction graph, Curr. Opin. Chem. Eng. 21 (2018), 48--59. \url{https://doi.org/10.1016/j.coche.2018.02.009}

\bibitem{Ottinger200201}H.C. \"{O}ttinger, M.A. Peletier,  A.  Montefusco,  A framework of nonequilibrium statistical mechanics. I. Role and type of fluctuations. arXiv preprint arXiv:2004.09120 (2020). \url{https://arxiv.org/abs/2004.09120}

\bibitem{Ottinger200202}A. Montefusco, M.A. Peletier, H. C.  \"{O}ttinger,  A framework of nonequilibrium statistical mechanics. II. Coarse-graining. arXiv preprint arXiv:2004.09121  (2020). \url{https://arxiv.org/abs/2004.09121}

\bibitem{Chew1961} G.F. Chew, The $S$-matrix theory of strong interactions.  (1961). Lawrence Berkeley National Laboratory, Report UCRL-9701, May 15, 1961.
\url{https://escholarship.org/content/qt2v38k2hr/qt2v38k2hr.pdf}

\bibitem{Chinesta2019}F. Chinesta, E. Cueto, M. Grmela, B.  Moya,  M.  Pavelka,  Learning physics from data: a thermodynamic interpretation. arXiv preprint arXiv:1909.01074 (2019). \url{https://arxiv.org/abs/1909.01074}

\bibitem{YBGE1991}G.S. Yablonskii, V.I. Bykov, A.N. Gorban,  V.I. Elokhin, Kinetic models of catalytic reactions. Elsevier, Amsterdam -- New York, 1991.

\end{thebibliography}
\end{document}